\documentclass[12pt]{article}
\usepackage{}
\usepackage{epsfig}

\usepackage{a4}
\usepackage{amsmath}

\newcommand{\insertplot}[5]{\begin{figure}
 \hfill\hbox to 0.05in{\vbox to #5in{\vfill
 \inputplot{#1}{#4}{#5}}\hfill}
 \hfill\vspace{-.1in}
 \caption{#2}\label{#3}
 \end{figure}}
 \newcommand{\inputplot}[3]{
 \special{ps: plotfile #1}
}
\newcounter{fig}   

\usepackage{epsfig}
\usepackage{amsmath}
\usepackage{amsfonts} 
\usepackage{graphicx}
\usepackage[german, english]{babel}
\usepackage{a4wide}
\usepackage{amsmath}
\usepackage{amssymb}
\usepackage{ifthen}
\usepackage{epsfig}

\begin{document}
\title{Abelian Hopfions of the $\mathbb C \mathbb P^n$ model on $\mathbb R^{2n+1}$
and a fractionally powered topological lower bound}
\author{{\large Eugen Radu,}$^{\star}$
{\large D. H. Tchrakian}$^{\dagger\ddagger}$
and {\large Yisong
Yang}$^{\sharp\diamond}$ \\ \\
$^{\star}${\small Institut f\"ur Physik, Universit\"at Oldenburg,
D-26111 Oldenburg, Germany}\\ \\
$^{\dagger}${\small School of Theoretical Physics -- DIAS, 10 Burlington
Road, Dublin 4, Ireland }\\
$^{\ddagger}${\small Department of Computer Science, NUIM, Maynooth,
Co. Kildare, Ireland}\\ \\
$^{\sharp}${\small Institute of Contemporary Mathematics, Henan University, Kaifeng, 
Henan 475004, PR China}\\
$^{\diamond}${\small Department of Mathematics, Polytechnic Institute of New York University,}
\\
{\small Brooklyn, New York 11201, USA}\\ \\
}

\date{}
\newcommand{\dd}{\mbox{d}}
\newcommand{\tr}{\mbox{tr}}
\newcommand{\nn}{\nonumber}
\newcommand{\la}{\lambda}
\newcommand{\ka}{\kappa}
\newcommand{\z}{\zeta}
\newcommand{\ta}{\theta}
\newcommand{\f}{\phi}
\newcommand{\vf}{\varphi}
\newcommand{\F}{\Phi}
\newcommand{\al}{\alpha}
\newcommand{\ga}{\gamma}
\newcommand{\de}{\delta}
\newcommand{\si}{\sigma}
\newcommand{\Si}{\Sigma}
\newcommand{\bomega}{\mbox{\boldmath $\omega$}}
\newcommand{\bsi}{\mbox{\boldmath $\sigma$}}
\newcommand{\bchi}{\mbox{\boldmath $\chi$}}
\newcommand{\bal}{\mbox{\boldmath $\alpha$}}
\newcommand{\bpsi}{\mbox{\boldmath $\psi$}}
\newcommand{\brho}{\mbox{\boldmath $\varrho$}}
\newcommand{\beps}{\mbox{\boldmath $\varepsilon$}}
\newcommand{\bxi}{\mbox{\boldmath $\xi$}}
\newcommand{\bbeta}{\mbox{\boldmath $\beta$}}
\newcommand{\ee}{\end{equation}}
\newcommand{\eea}{\end{eqnarray}}
\newcommand{\be}{\begin{equation}}
\newcommand{\bea}{\begin{eqnarray}}

\newcommand{\ii}{\mbox{i}}
\newcommand{\e}{\mbox{e}}
\newcommand{\pa}{\partial}
\newcommand{\Om}{\Omega}
\newcommand{\vep}{\varepsilon}
\newcommand{\bfph}{{\bf \phi}}
\newcommand{\lm}{\lambda}
\def\theequation{\arabic{equation}}
\renewcommand{\thefootnote}{\fnsymbol{footnote}}
\newcommand{\re}[1]{(\ref{#1})}

\newcommand{\R}{{\rm I \hspace{-0.52ex} R}}
\newcommand{\N}{{\sf N\hspace*{-1.0ex}\rule{0.15ex}%
{1.3ex}\hspace*{1.0ex}}}
\newcommand{\Q}{{\sf Q\hspace*{-1.1ex}\rule{0.15ex}%
{1.5ex}\hspace*{1.1ex}}}
\newcommand{\C}{{\sf C\hspace*{-0.9ex}\rule{0.15ex}%
{1.3ex}\hspace*{0.9ex}}}
\newcommand{\p}{{\sf P\hspace*{-0.9ex}\rule{0.15ex}%
{1.3ex}\hspace*{0.9ex}}}
\newcommand{\eins}{1\hspace{-0.56ex}{\rm I}}
\renewcommand{\thefootnote}{\arabic{footnote}}

\maketitle


\bigskip

\begin{abstract}
Regarding the Skyrme-Faddeev model on $\mathbb R^3$ as a $\mathbb C \mathbb P^1$ sigma model, 
we propose  $\mathbb C \mathbb P^n$ sigma models on $\mathbb R^{2n+1}$
as generalisations which may support finite energy Hopfion solutions in these dimensions. 
The topological charge stabilising these field configurations
is the Chern-Simons charge, namely the volume integral of the Chern-Simons density 
which has a local expression in terms of the composite connection
and curvature of the $\mathbb C \mathbb P^n$ field. 
It turns out that subject to the sigma model constraint,
this density is a total divergence. 
We  prove the existence of a topological lower bound on the energy,
which, as in the Vakulenko-Kapitansky case in $\mathbb R^3$,
is a fractional power of the topological charge, depending on $n$. 
The numerical construction of the simplest ring shaped un-knot Hopfion on $\mathbb R^5$
is also discussed. 
\end{abstract}
\medskip
\medskip

\section{Introduction}
Hopfions are knotted, static, finite energy and topologically stable solutions to nonlinear differential equations.
To date, the only known model supporting Hopfions is the Skyrme-Faddeev model~\cite{F}, which is a $O(3)$ sigma model on $\mathbb R^3$.
These Hopf soliton solutions are constructed numerically~\cite{Faddeev:1996zj,Battye:1998pe,Radu:2008pp} and the topological charge stabilising
them is the volume integral of the Chern--Simons (CS) density~\cite{Manton:2004tk}. This, namely being stabilised by the CS charge, is the distinguising feature of the
known~\cite{F} Hopfions on $\mathbb R^3$, as well as the higher dimensional ones to be introduced below.

In the definition of a Chern-Simons charge, the entities empolyed are the gauge field connection and curvature. But if these entities be the Yang-Mills
connection and curvature, then the resulting Chern-Simons density is manifestly {\bf not} a total divergence, and hence is {\bf not} a candidate for
a topological charge density. What confers the status of a topolological charge density on the CS density is, that the latter be defined in terms of
the composite gauge field connection and curvature of a suitable nonlinear sigma model described by a scalar field. In this case, it is possible to show
that such a CS density is indeed total divergence~\footnote{In the familiar case of the Skyrme-Faddeev model, the CS density
is $nonlocal$ and hence cannot be expressed in this way, as it stands. However, since the $O(3)$ sigma model is equivalent to the $\mathbb C \mathbb P^1$ model,
this assertion remains true.}, which means that the volume integral of the CS density can be evaluated as a surface integral,
rendering it a candidate for a topological charge density. 
This is done by subjecting the CS density to the variational principle
taking the sigma model constraint into account. The resulting field equations turn out to be trivial, encoding no dynamics.
We have described such topological charge densities as ``essentially total derivative'' below~\footnote{The Chern-Simons densities of sigma models supporting
Hopfions share this property with the topological charge densities of the $O(D+1)$ sigma models on $\mathbb R^D$.}.
The role of the Chern--Simons (CS) density as the topological charge density, is the distinguishing feature of field theories that support Hopf
solitons~\footnote{In contrast, the solitons of the
Grassmannian and Yang-Mills models in all even dimensions~\cite{Tchrakian:2010ar} are stabilised by Chern-Pontryagin (CP) charges, and the solitons of the
Yang-Mills-Higgs models in all dimensions~\cite{Tchrakian:2010ar} by the descendents of CP charges.}.

Concerning the role of composite connections in the definition of the CS density, it is clear that these can be either Abelian or non-Abelian.
As suggested by the title here, the present work is concerned exclusively with the case of Abelian composite connections.



A very important difference between the usual solitons and Hopfions is, that in the former the topological lower bound on the energy is
established by the familiar Bogomol'nyi-type procedure employing the Chern-Pontryagin charge density (or its descendents). In the case of Hopfions
on the other hand, this lower bound is established by the considerably more elaborate procedure of Vakulenko and
Kapitansky~\cite{Vakulenko:1979uw,Lin-Yang1}, hinged instead on the Chern-Simons density, and employing a Sobolev space technique.
The resulting topological lower bounds are expressed as fractional powers of the topolocical charge, namely the Chern-Simons charge.
This type of lower bound is essential in forming knotted solitons.

Finally we note, in passing, that the CS densities used in the context of Hopfions
are defined on the odd dimensional Euclidean spaces $\mathbb R^{2n+1}$, and not on Minkowskian spacetimes in those dimensions.
They encode only geometric and topological data
and do not enter the Lagrangian in Minkowski space as in Ref~\cite{Deser:1982vy}. As such, they are entirely different from the $dynamical$
Chern-Simons terms appearing in field theories introduced in Ref~\cite{Deser:1982vy}.

The aim of this work is to propose a family of sigma models that we claim can support Hopfion solutions\footnote{Higher dimensional
models supporting Hopfions were considered earlier~\cite{Lin-Yang12,Lin-Yang2}, but these are $O(2n+1)$ sigma models rather than the $\mathbb C \mathbb P^n$ models proposed
here. In contrast to the former, the CS density of the latter has a local expression.} on $\mathbb R^{2n+1}$, the $n=1$ case being the well-known
Skyrme-Faddeev Hopfion. We have referred to these as Abelian Hopfions since the composite connections of the sigma models in question, namely the
$\mathbb C \mathbb P^n$ models, are Abelian. The $\mathbb C \mathbb P^n$ field systems are presented in section {\bf 2}, where the general properties of the
Chern-Simons densities are also discussed.  In section {\bf 3} we propose models that can support finite energy Hopfion solutions,
with special regard to section {\bf 4} following it, where the energy lower bounds for some of these models is established.
In section {\bf 5} we deal with the detailed properties of the Chern-Simons charges, subjecting them to suitable symmetries that result in
revealing the boundary values required of Hopfion field configurations. 
Some numerical results for the simplest Hopfions in  $D=5$ dimensions are reported in section {\bf 6}.
Section {\bf 7} is devoted to a summary and a discussion of our results.

\section{The $\mathbb C \mathbb P^n$ fields on $\mathbb R^{2n+1}$}
We start with the generic structure of models that can support Abelian Hopfion on
$\mathbb R^{2n+1}$. These are described by complex $(n+1)-$tuplets
\be
\label{field}
Z=\left[
\begin{array}{c}
z_1\\
z_2\\
..\\
..\\
z_{n+1}
\end{array}
\right]\equiv z_a\ ;\ \ \bar Z=\left[
\begin{array}{c}
\bar z^1\\
\bar z^2\\
..\\
..\\
\bar z^{n+1}
\end{array}
\right]\equiv \bar z^a\ ,\quad a=1,2,...,n+1\,,
\ee
subject to the constraint
\be
\label{constr}
Z^{\dagger}\,Z\equiv\bar z^a\,z_a=1\,,
\ee
taking their values in
$\frac{U(n+1)}{U(n)\times U(1)}$, parametrised by $2n$ real functions. In \re{field}, $\bar z^a$ is the complex conjugate of $z_a$, transforming with an index
that is $contravariant$ to the $covariant$ index of $z_a$, and $Z^{\dagger}$ in \re{constr} is the transpose of $\bar Z$.
This leads to the definition of the $projection$ $operator$
\be
\label{proj}
P=\left(\eins-Z\,Z^{\dagger}\right)\equiv\left(\delta_a{}^b-z_a\,\bar z^b\right)\,.
\ee
The most interesting feature of these models is that when the field $Z$ is subjected to
a {\it local} $U(1)$ gauge transformation $g=\e^{\mp\,i\Lambda(x)}$, under which the constraint \re{constr} is invariant, then the quantity defined as
\be
\label{comp}
B_{i}=i\,Z^{\dagger}\pa_{i}Z\ ,\quad i=1,2,...,2n+1\ ,
\ee
transforms like an {\it Abelian composite connection} under $g(\Lambda)$,
\[
B_{i}\to B_i\pm\pa_i\Lambda\,.
\]
This leads to the
definition of the covariant derivative of $Z$ and and the {\it Abelian curvature} of this connection,
\bea
D_iZ&=&\pa_i\,Z+B_i\,Z\,,
\label{cov1}
\\
G_{ij}&=&\pa_i\,B_j-\pa_j\,B_i\,,\label{curv1}
\eea
with $D_iZ$ transforming covariantly under the action of $g$, and with $G_{ij}$ invariant.

The Abelian Chern-Simons (CS) density on $\mathbb R^{2n+1}$ is then readily defined in terms of the quantities
\re{curv1} and \re{cov1},
\be
\label{CS}
\Omega_{\rm CS}\simeq\vep_{i_1i_2\dots i_{2n+1}}\,B_{i_{2n+1}}\,G_{i_1i_2}\,G_{i_3i_4}\dots G_{i_{2n-1}i_{2n}}\,.
\ee
When it is subjected to arbitrary variations of $Z^{\dagger}$ (or $Z$),
with adequate account taken of the constraint \re{constr}, this yields the variational equation
\be
\label{eqn}
\vep_{i_1i_2\dots i_{2n+1}}\,D_{i_{2n+1}}Z\,G_{i_1i_2}\,G_{i_3i_4}\dots G_{i_{2n-1}i_{2n}}=0~,
\ee
which on the face of it is a nontrivial equation of motion for $Z$, meaning that the CS density $\Omega_{\rm CS}$ is not a total divergence. This however is not true.
It can be verified by direct calculation, with the constraint \re{constr} taken account of,
that $\Omega_{\rm CS}$ is indeed a total divergence and that \re{eqn} is actually trivial~\footnote{It should be remarked here that in the case of
the $O(D+1)$ sigma models on $\mathbb R^{D}$, the situation is much simpler, with the analogue of Eqn. \re{eqn} displaying a vanishing left hand side.} and encodes no
dynamical information.

\section{The $\mathbb C \mathbb P^n$ ``Skyrme-Faddeev'' models on $\mathbb R^{2n+1}$}
We refer to the models supporting Hopfions as ``Skyrme-Faddeev'' models, to distinguish them from the ``Skyrme''
models that support usual Skyrmions.
Both are sigma models, for example the Skyrme-Faddeev model and the Skyrme model on $\mathbb R^{3}$ are both $O(3)$ sigma models. 
These however differ from each other
in that the highest order kinetic term in the former is the quartic Skyrme term, while that in the latter is the sextic Skyrme term. These are respectively
the squares of a $2$-form and a $3$-form field constructed from the partial derivatives af the scalar sigma model field. The highest order form that can be constructed
is limited by the number of independent real functions parametrising the sigma model field. In the case of the $\mathbb C \mathbb P^n$ fields here, this number is $2n$ so that
the highest order kinetic term in the model supporting a Hopfion on $\mathbb R^{2n+1}$ is the square of a $2n$-, or $(D-1)$-form field. This contrasts with the situation of,
for example, the (usual) Skyrmions of the $O(D+1)$ sigma models on $\mathbb R^{D}$. In that case the number of independent functions is $D$, so that
the highest order kinetic term is the square of a $D$-form field.


Irrespective of what symmetries the models are subjected to, their energy density functionals must be consistent with Derrick's scaling requirement.
This is so that the energy of the Hopfion solutions be finite. In this respect, the situation is the same as that for the usual Skyrme models,
except for the fact that the order of the highest order kinetic term for the Skyrme-Faddeev models is of lower order as explained in the previous paragraph.

The knotted solitons supported by these systems are topologically stable and their energies are bounded from below by $n$-dependent fractional powers
of the Chern-Simons charge.

\subsection{The $\mathbb C \mathbb P^1$ ``Skyrme-Faddeev'' models on $\mathbb R^3$}
The most general model supporting finite energy solutions, consistent with the Derrick scaling requirement is
\be
\label{model13}
{\cal E}_3=\ka_0^0\,V+\frac12\,\ka_1^2\,D_iZ^{\dagger}\,D_iZ
+\frac14\,\ka_2^4\,G_{ij}^2~,
\ee
with $D_iZ$ and $G_{ij}$ given by \re{cov1} and \re{curv1}. The constants
$\ka_0$, $\ka_1$, and $\ka_2$ each have the dimension of length, and $V$ is some pion
mass type potential, which can most naturally be chosen to be
\be
\label{pot}
V=1+Z^{\dagger}\,\si_3\,Z\,.
\ee
It is well known that a $\mathbb C \mathbb P^1$ sigma model is equivalent~\footnote{The $O(3)$ field $\f^a$, $a=1,2,3$, subject to the constraint $|\f^a|^2=1$ is
given by$$\f^a=Z^{\dagger}\,\si^a\,Z$$in terms of the Pauli matrices $\si^a$ and the field $Z$ in \re{field} for $n=1$.} to the corresponding $O(3)$ sigma model.
In the present case, the $\mathbb C \mathbb P^1$ system \re{model13} with $\ka_0=0$, is equivalent to the $O(3)$ Skyrme-Faddeev model. The extension of the latter
with the potential term \re{pot} has been studied recently in \cite{Foster:2010zb}, where Hopfion solutions are constructed.

The virial identity resulting from the scaling requirement that must be satisfied is
\be
\label{virial1}
3\|V\|+\|D_iZ\|^2-2\|G_{ij}\|^2=0\,,
\ee
where the dimensional constants and the detailed normalisations have been suppressed, and where each of the
quantities $\|.\|^2$ is the positive definite integral of the correspoding density in \re{model13}.

The lower bound on the energy of \re{model13} has been established long ago in \cite{Vakulenko:1979uw}. The presence of the potential term \re{pot} does
not influence this lower bound.

\subsection{The $\mathbb C \mathbb P^2$ ``Skyrme-Faddeev'' models on $\mathbb R^5$}
The most general model supporting finite energy solutions in this dimension, consistent with the Derrick scaling requirement is
\be
\label{model15}
{\cal E}_5=\ka_0^0\,V+\frac12\,\ka_1^2\,D_iZ^{\dagger}\,D_iZ
+\frac14\,\ka_2^4\,G_{ij}^2
+\frac18\,\ka_3^6\,(G_{[ij}\,D_{k]}Z)^{\dagger}(G_{[ij}\,D_{k]}Z)
+\frac{1}{16}\,\ka_4^8\,G_{ijkl}^2~,
\ee
with the $4-$form (composite) curvature $G_{ijkl}$ being
the totally antisymmetrised two-fold product of the $2-$form (composite) curvature $G_{ij}$, and the square brackets on the indices implying cyclic symmetry. The constants
$\ka_0$, $\ka_1$, $\ka_2$, $\ka_3$ and $\ka_4$ each have the dimension of length, and $V$ is some pion
mass type potential. According to the scaling requirement for finite energy, it is necessary to retain {\it at least one}
of the constants $(\ka_1,\ka_2)$ and {\it at least one} of the constants $(\ka_3,\ka_4)$, with the option of
setting the rest equal to zero.

The virial identity resulting from the scaling requirement that must be satisfied is
\be\label{virial2}
5\|V\|+3\|D_iZ\|^2
+\|G_{ij}\|^2
-\|(G_{[ij}\,D_{k]}Z)\|^2
-3\|G_{ijkl}\|^2=0\,,
\ee
where the dimensional constants and the detailed normalisations have been suppressed, and where each of the
quantities $\|.\|^2$ is the positive definite integral of the correspoding density in \re{model15}.

\subsection{The $\mathbb C \mathbb P^3$ ``Skyrme-Faddeev'' models on $\mathbb R^7$}

The most general model supporting finite energy solutions here, consistent with the Derrick scaling requirement is
\bea
\label{model17}
{\cal E}_7=\ka_0^0\,V&+&\frac12\,\ka_1^2\,D_iZ^{\dagger}\,D_iZ
+\frac14\,\ka_2^4\,G_{ij}^2
+\frac18\,\ka_3^6\,(G_{[ij}\,D_{k]}Z)^{\dagger}(G_{[ij}\,D_{k]}Z)\nonumber\\
&&+\frac{1}{16}\,\ka_4^8\,G_{ijkl}^2
+\frac{1}{32}\,\ka_5^{10}\,(G_{[ijkl}\,D_{m]}Z)^{\dagger}(G_{[ijkl}\,D_{m]}Z)
+\frac{1}{32}\,\ka_5^{12}\,G_{ijklmn}^2
\eea
with the $6-$form (composite) curvature $G_{ijklmn}$ being
the totally antisymmetrised three-fold product of the $2-$form (composite) curvature $G_{ij}$, and the square brackets on the indices again implying total
antisymmetrisation. The constants
$\ka_0$, $\ka_1$, $\ka_2$, $\ka_3$, $\ka_4$, $\ka_5$ and $\ka_6$ each have the dimension of length, and $V$ is some pion
mass type potential. According to the scaling requirement for finite energy, it is necessary to retain {\it at least one}
of the constants $(\ka_1,\ka_2,\ka_3)$ and {\it at least one} of the constants $(\ka_4,\ka_5,\ka_6)$, with the option of
setting the rest equal to zero.

The virial identity resulting from the scaling requirement that must be satisfied is
\be
\label{virial3}
7\|V\|+5\|D_iZ\|^2+3\|G_{ij}\|^2+\|(G_{[ij}\,D_{k]}Z)\|^2
-\|G_{ijkl}\|^2-3\|(G_{[ijkl}\,D_{m]}Z)\|^2-5\|G_{ijklmn}\|^2=0\,,
\ee
where the dimensional constants and the detailed normalisations have been suppressed, and where each of the
quantities $\|.\|^2$ is the positive definite integral of the correspoding density in \re{model17}.

\subsection{The general case: $\mathbb C \mathbb P^n$ models on $\mathbb R^{2n+1}$}
It is convenient here to introduce a dedicated notation, whereby the the $p$-form fields and curvatures are denoted by
\be
\label{notation}
G(0)=V\ , \ G(1)=D_iZ\ , \ G(2)=G_{ij}\ , \  G(3)=G(2)\wedge{G(1)}\,,\dots, \ G(2n-1)\ , \  G(2n)
\ee
such that the energy density functional can be written as
\be
\label{model12n+1}
{\cal E}_{2n+1}=G(0)+|G(1)|^2+G(2)^2+|G(3)|^2+\dots+|G(2n-1)|^2+G(2n)^2~.
\ee
The virial identity in this case is
\bea
\label{virial3-n}
&&(2n+1)\|G(0)\|+(2n-1)\|G(1)\|^2+(2n-3)\|G(2)\|^2+\dots+\|G(2n-3)\|^2-\nonumber\\
&&\quad\quad-\|G(2n-4)\|^2-\dots-
(2n-3)\|G(2n-1)\|^2-(2n-1)\|G(2n)\|^2=0\,.
\eea
Anticipating our considerations in the next section, where a topological bound on the energy is established, we point out that in all but the $n=1$ case
the most general models proposed feature the two terms constructed from the composite curvature $2$-form $G(2)$ and the $2n$-form composite curvature $G(2n)$.
The lower bound established is for the systems consisting of these two kinetic terms, and remains 
valid when the other positive definite terms are added.

\section{A topological lower bound}
In $2n+1$ dimensions, use $G_{i_1 i_2\cdots i_{2n}}$ to denote the $2n$-form curvature which is formed from taking the totally antisymmetric $n$-fold product
of the $2$-form curvature $G_{ij}$. Then
\be
\Omega_{\mbox{\small CS}}=\epsilon^{ij_1 j_2\cdots j_{2n}} B_i G_{j_1 j_2\cdots j_{2n}}~,
\ee
so that
\be
Q=\int\Omega_{\mbox{\small CS}}
\ee
is the topological charge, where the integral is understood to be evaluated over the domain space $\mathbb R^{2n+1}$.
The construction of $G_{i_1 i_2\cdots i_{2n}}$ leads naturally to the point-wise bound
\be \label{GG}
|G_{i_1 i_2\cdots i_{2n}}|\leq C|G_{ij}|^n,
\ee
where and in the sequel we use $C$ to denote a generic positive constant.

Consider the energy functional
\be \label{EG}
E=\int (G^2_{ij}+G^2_{i_1 i_2\cdots i_{2n}})
\ee

We shall establish the following fractional-exponent topological lower bound
\be \label{EB}
E\geq C|Q|^{\frac{2n+1}{2n+2}},\quad n\in\mathbb N.
\ee

To prove this lower bound, we first apply the Cauchy--Schwarz inequality to get
\be \label{Q1}
|Q|\leq C\left(\int |B_i|^p\right)^{\frac1p}\left(\int|G_{j_1 j_2\cdots j_{2n}}|^q\right)^{\frac1q},
\ee
where $p,q>1$ satisfies
\be\label{pq}
\frac1p+\frac1q=1.
\ee
Next, we recall the Sobolev inequality over $\mathbb R^m$:
\be \label{So1}
\|f\|_p\leq C\|\nabla f\|_2,
\ee
where we use $\|\cdot\|_p$ to denote the usual $L^p$-norm for a function defined over the space $\mathbb R^m$ and $p$ satisfies
the relation
\be \label{x56}
\frac1p=\frac12-\frac1m=\frac{m-2}{2m}.
\ee

Setting $m=2n+1$ in (\ref{x56}), we see that $p$ is given by
\be \label{p}
p=\frac{2(2n+1)}{2n-1}.
\ee
Inserting (\ref{p}) into (\ref{So1}), we have
\be \label{So2}
\|f\|_{\frac{2(2n+1)}{2n-1}}\leq C\|\nabla f\|_2.
\ee
Moreover, restricting to the Coulomb gauge, $\partial_i B_i=0$, there holds
\be \label{GB}
\int G_{ij}^2=\int(\partial_i B_j-\partial_j B_i)^2=\int|\nabla B_i|^2.
\ee
In view of (\ref{So2}) and (\ref{GB}), we obtain
\be \label{Bi}
\left(\int |B_i|^{\frac{2(2n+1)}{2n-1}}\right)^{\frac{2n-1}{2(2n+1)}}\leq C\left(\int G_{ij}^2\right)^{\frac12}.
\ee

On the other hand, in view of (\ref{pq}) and  (\ref{p}) we see that
\be
q=\frac{2(2n+1)}{2n+3},
\ee
and in view of (\ref{Q1}) and (\ref{Bi}) we conclude that
the bound
\be \label{Q2}
|Q|\leq C\left(\int G_{ij}^2\right)^{\frac12}\left(\int |G_{j_1 j_2\cdots j_{2n}}|^{\frac{2(2n+1)}{2n+3}}\right)^{\frac{2n+3}{2(2n+1)}}
\ee
is valid.

To proceed further, we seek constants $\alpha,\beta>0$ and $s,t>1$ such that
\be \label{ab}
\left.\begin{array}{rll}\alpha+\beta&=&\frac{2(2n+1)}{2n+3},\\ \frac1s+\frac1t&=&1,\\ \alpha s&=&\frac2n,\\
\beta t&=&2.\end{array}\right\}
\ee
If the system (\ref{ab}) has a solution, then we can use the Cauchy--Schwarz inequality to reduce the second integral on the right-hand side of
(\ref{Q2}) into
\bea \label{G3}
\int |G_{j_1 j_2\cdots j_{2n}}|^{\frac{2(2n+1)}{2n+3}}&=&\int|G_{j_1 j_2\cdots j_{2n}}|^{\alpha+\beta}\nn\\
&\leq&\left(\int|G_{j_1j_2\cdots j_{2n}}|^{\alpha s}\right)^{\frac1s}\left(\int |G_{j_1j_2\cdots j_{2n}}|^{\beta t}\right)^{\frac1t}\nn\\
&=&\left(\int|G_{j_1j_2\cdots j_{2n}}|^{\frac2n}\right)^{\frac1s}\left(\int |G_{j_1j_2\cdots j_{2n}}|^{2}\right)^{\frac1t}.
\eea
Fortunately, it is not hard to see that when $n\neq1$ the system (\ref{ab}) may be uniquely solved to yield
\be \label{ab2}
\left.\begin{array}{rll}
\alpha &=&\frac4{(n-1)(2n+3)},\\ \beta &=&\frac{2(2n^2-n-3)}{(n-1)(2n+3)},\\ s &=&\frac{(n-1)(2n+3)}{2n},\\ 
t&=&\frac{(n-1)(2n+3)}{2n^2-n-3}.\end{array}\right\}
\ee
Substituting (\ref{ab2}) into (\ref{G3}) and applying (\ref{GG}), we arrive at
\be \label{G4}
\int |G_{j_1 j_2\cdots j_{2n}}|^{\frac{2(2n+1)}{2n+3}}\leq \left(\int|G_{ij}|^2\right)^{\frac{2n}{(n-1)(2n+3)}}\left(\int |G_{j_1j_2\cdots j_{2n}}|^{2}\right)^{\frac{2n^2-n-3}{(n-1)(2n+3)}}.
\ee
Now inserting (\ref{G4}) into (\ref{Q2}) and applying (\ref{EG}), we get
\bea\label{Q3}
|Q| &\leq& C\left(\int G_{ij}^2\right)^{\frac12+\frac{n}{(n-1)(2n+1)}}\left(\int |G_{j_1 j_2\cdots j_{2n}}|^2\right)^{\frac{2n^2-n-3}{2(n-1)(2n+1)}}\nn\\
&\leq& CE^{\frac12+\frac{n}{(n-1)(2n+1)}+\frac{2n^2-n-3}{2(n-1)(2n+1)}}\nn\\
&=& CE^{\frac{2(n+1)}{2n+1}},\quad n\neq1.
\eea

 In other words, the energy-topological charge inequality (\ref{EB}) is proved when $n\neq1$. 
However, when $n=1$,  the classical Vakulenko--Kapitansky
inequality holds.
Thus we see that (\ref{EB}) is valid for all $n\in{\mathbb N}$.

\section{Imposition of symmetries on $\mathbb C \mathbb P^n$ models on $\mathbb R^{2n+1}$}
The imposition of symmetries on a system obeying partial differential eaquations (PDE) serves the purpose of lowering the order of the PDE's. This
is useful in the analytic proof of existence of some solutions, and is often necessary in the numerical construction of solutions. 
Indeed in the last section, we
present some numerical results for a set of second order PDE's
with dependence of two variables only,
which result from a suitable imposition of symmetry.

There is however another reason for the imposition of symmetries, relating to the evaluation of the topological charge.
We know that the $\mathbb C \mathbb P^n$ field $Z$ on $\mathbb R^{2n+1}$ is parametrised by $2n$ real functions, and since the Chern-Simons density is (effectively) a 
total divergence, then the number of effective space coordinates for the field configuration describing a topologically stable Hopfion
must also be $2n$. The effective space coordinates must reflect the symmetry of the Hopfion in question. In general, this symmetry can be quite complicated and the
resulting evaluation of the integral giving the topological charge may not be transparent. It is our aim here to introduce a symmetry imposition criterion that
renders the statement of the boundary values of the Hopfion field configuration consistent with integer topological charge, transparent. 


The topological charge, which is the volume integral of the CS density, is evaluated as a surface integral. Our criterion in the imposition of symmetries is that
in its final form, after the imposition of symmetry, the CS density be expressed as a totally antisymmetrised product of the residual space derivatives of the residual
array of functions parametrising the system, $i.e.$ that it be expressed as a total divergence explicitly. 

 As $per$ this criterion, the minimal imposition of symmetry is axial symmetry in any one of the $2$-planes in $\mathbb R^{2n+1}$. This results in decreasing the number
of residual space coordinates to $2n$, that being the number of independent real functions parametrising the $\mathbb C \mathbb P^n$ field $Z$ (and hence $B_i$ and the CS
density), namely $2n$. We refer to this as the minimal imposition of symmetry. 

The maximal imposition of symmetry, consistent with our criterion
is that of the imposition of azimuthal symmetry in each of the $n$, $2$-planes in $\mathbb R^{2n+1}$, $i.e.$ $n$-fold azimuthal symmetry.

It is clear that in between the minimal, namely mono-azimuthal or axial symmetry, and the maximal $n$-fold azimuthal symmetry, one can impose
any number $N (< n)$ of azimuthal symmetries. In the following, we shall ingore these and concentrate only the maximal and the minimal cases.

\subsection{Imposition of (minimal) mono-azimuthal symmetry}
As explained, the function $Z$ must be subjected to at least one azimuthal symmetry, and this is the case considered here. (We ignore intermediate cases
where more than one but less than $n$ azimuthal symmetries can be imposed.)

We will apply axial symmetry to only one $2$-plane in $\mathbb R^{2n+1}$, say to the $x_{\al}=(x_{2n},x_{2n+1})$ plane. We denote the radial coordinate in this plane
by $\z=\sqrt{|x_{\al}|^2}$, the azimuthal angle by $\vf$ and the vortex number by $n$. The coordinate on $\mathbb R^{2n+1}$ is denoted by $x_i$, with $i=1,2,\dots,(2n+1)$.
Denoting the coordinates in the plane where axial symmetry is imposed by $x_{\al}$, we split the index $i$ as
\[
x_i=(x_a,x_{\al})\ , \quad\al=2n,2n+1\ ,\quad a=1,2,\dots,(2n-1)\,.
\]
The mono-azimuthally symmetric Ansatz for $Z$ on $\mathbb R^{2n+1}$ is
\be
\label{genaz}
Z=\left[
\begin{array}{c}
a_1+ib_1\\
a_2+ib_2\\
a_3+ib_3\\
\dots\\
a_n+ib_n\\
c\,\e^{in\vf}
\end{array}
\right]\,,
\ee
which subject to the constraint \re{constr} satisfies
\be
\label{constrmono}
(a_1^2+a_2^2+\dots+a_n^2)+(b_1^2+b_2^2+\dots+b_n^2)+c^2=1\,,
\ee
resulting in the maximal number of $2n$ independent real functions parametrizing $Z$, each of these depending on the $2n$ vaiables $(x_a,\z)$, $a=1,2,\dots,2n-1$.

Substitution of \re{genaz} in the Abelian CS density \re{CS} yields a result of the form
\be
\label{redCSgen}
\Omega_{\rm CS}\simeq\frac{n}{\z}\,c\,\cdot\mbox{det}
\left|
\begin{array}{cccccc}
a_{1}\ \ \ \ \ \ \ \ b_{1}\ \ \ \ \ \ \ a_{2}\ \ \ \ \ \ \ b_{2}\ \  \dots\ a_{n}\ \ \ \ \ \ \ b_{n}\ \ \ \ \ \ \ c\\
a_{1,1}\ \ \ \ \ \ \  b_{1,1}\ \ \ \ \ a_{2,1}\ \ \ \ b_{2,1}\ \dots\ \ a_{n,1}\ \ \ \ \ b_{n,1}\ \ \ \ \ \ c_{,1}\\
a_{1,2}\ \ \ \ \ \ b_{1,2}\ \ \ \ \ \ a_{2,2}\ \ \ \ \ b_{2,2} \ \ \ \dots\ a_{n,2}\ \ \ \ \ \ b_{n,2}\ \ \ \ \ c_{,2}\\
a_{1,3}\ \ \ \ \ \ b_{1,3}\ \ \ \ \ \ a_{2,3}\  \ \ \ \ b_{2,3}\ \dots\ a_{n,3}\ \ \ \ \ \ b_{n,3}\ \ \ \ \ c_{,3}\\
\dots\\
a_{1,2n-1}\ b_{1,2n-1}\ a_{2,2n-1}\ b_{2,2n-1}\ \dots\ a_{n,2n-1}\ b_{n,2n-1}\ c_{,2n-1}\\
\ a_{1,\z}\ \ \ \ \ \ \, b_{1,\z}\ \ \ \ \ \, a_{2,\z}\ \ \ \ \ \, b_{2,\z}\ \dots\ \ \, a_{n,\z}\ \ \ \ \ \, b_{n,\z}\ \ \ \ \  c_{\z}
\end{array}
\right|\,,
\ee
where $a_{1,\z}=\pa_{\z}a_1$, $a_{1,a}=\pa_{x^a} a_1$, with $a=1,2,\dots,2n-1$, $etc.$

The determinant \re{redCSgen} is not explicitly a total divergence but if subjected to the variational principle taking account of the constraint
\re{constrmono}, yields no nontrivial equation of motion, $i.e.$ it is ``essentially total divergence''.

Of course, if \re{genaz} is reparametrised in a convenient polar
parametrisation satisfying \re{constrmono}, then the reduced CS density will become  ``explicitly total divergence''. It is worthwhile giving some
examples of such polar parametrisations, to help illustrate the boundary values that should be satisfied for the (topological) charge integral to take integer
values. To this end, consider the two examples with~\footnote{We omit the $n=1,D=3$ case since this is identical with the $n=1,D=3$ case considered in section
{\bf 5.2.1} next.} $n=2,D=5$ and $n=3,D=7$. There
\be
\label{zA5}
Z=\left[
\begin{array}{c}
a_1+ib_1\\
a_2+ib_2\\
c\,\e^{in\vf}
\end{array}
\right]
\equiv\left[
\begin{array}{c}
\sin\frac12f\,\sin g\,\e^{i\al}\\
\sin\frac12f\,\cos g\,\e^{i\beta}\\
\cos\frac12f\,\e^{in\vf}
\end{array}
\right]
\ee
and
\be
\label{zA7}
Z=\left[
\begin{array}{c}
a_1+ib_1\\
a_2+ib_2\\
a_3+ib_3\\
c\,\e^{in\vf}
\end{array}
\right]
=\left[
\begin{array}{c}
\sin\frac12f\,\sin g_1\,\cos g_2\,\e^{i\al}\\
\sin\frac12f\,\sin g_1\,\sin g_2\,\e^{i\beta}\\
\sin\frac12f\,\cos g_1\,\e^{i\ga}\\
\cos\frac12f\,\e^{in\vf}
\end{array}
\right]\ ,
\ee
respectively, which automatically satisfy the constraints \re{constrmono}. Each of the functions in \re{zA5} and \re{zA7} depends on the $2n$
vaiables $0\le\z\le\infty$ and $-\infty\le x_a\le\infty$, $a=1,2,\dots,2n-1$, for $n=2$ and $n=4$ respectively.

We define
\be
\label{xi4}
\xi_{\mu}=\left(
\begin{array}{c}
r\,\sin\ta\,\cos\vf_1\\
r\,\sin\ta\,\sin\vf_1\\
r\,\cos\ta\,\cos\vf_2\\
r\,\cos\ta\,\sin\vf_2
\end{array}\right)
\ee
with $0\le\vf_1\le 2\pi$, $0\le\vf_2\le 2\pi$ and $0\le\ta\le\frac{\pi}{2}$ for \re{zA5}, and,
\be
\label{xi5}
\xi_{\mu}=\left(
\begin{array}{c}
r\,\sin\ta_1\,\cos\ta_2\,\cos\vf_1\\
r\,\sin\ta_1\,\cos\ta_2\,\sin\vf_1\\
r\,\sin\ta_1\,\sin\ta_2\,\cos\vf_2\\
r\,\sin\ta_1\,\sin\ta_2\,\sin\vf_2\\
r\,\cos\ta_1\,\cos\vf_3\\
r\,\cos\ta_1\,\sin\vf_3
\end{array}\right)
\ee
with $0\le\vf_1\le 2\pi$, $0\le\vf_2\le 2\pi$, $0\le\vf_3\le 2\pi$, $0\le\ta_1\le\frac{\pi}{2}$ and $0\le\ta_2\le\frac{\pi}{2}$ for \re{zA7}.

The boundary values required of the functions in $(f,g,\al,\beta)$ in \re{zA5} and the functions $(f,g_1,g_2,\al,\beta,\ga)$ in \re{zA7}, such
that the topological charge be integer (with suitable normalisation) can be stated very naturally. As in the corresponding maximal symmetry
cases considered in sections {\bf 5.2.1}, {\bf 5.2.2}, and {\bf 5.2.3}, above, the function $f$ is assigned the asymptotic value
\[
\lim_{r\to\infty}f=0\,.
\]
The rest of the functions in \re{zA5}-\re{zA7} are then assigned, respectively
\bea
&&\lim_{r\to\infty}g=\ta\ ,\quad\lim_{r\to\infty}\al=m_1\,\vf_1\ ,\quad\lim_{r\to\infty}\beta=m_2\,\vf_2\,.\label{bv5}\\
&&\lim_{r\to\infty}g_1=\ta_1\, ,\  \lim_{r\to\infty}g_2=\ta_2
\ ,\  \lim_{r\to\infty}\al=m_1\,\vf_1\, ,\  \lim_{r\to\infty}\beta=m_2\,\vf_2\, , \ \lim_{r\to\infty}\ga=m_3\,\vf_3\,,\label{bv7}
\eea
the case of arbitrary dimension following by induction.

The Chern-Simons topological charges in the generic case will be of the form
\[
Q\simeq n\,m_1\,m_2\,\dots m_n\,\pi^{n+1}\,.
\]

\subsection{Imposition of (maximal) $n$-fold azimuthal symmetry}
In these cases the residual systems are parametrised by $(n+1)$-dimensional spatial coordinates. Then, the application of Gauss' Theorem on the volume
integral of the CS density, expressed as a total divergence, results in $n$ dimensional angular integrals yielding the CS charges.
This is carried out explicitly for $n=1,2,3$, and the generic case is deduced by induction.

\subsubsection{Chern-Simons charge on $\mathbb R^3$ subject to axial symmetry}
This is the usual Skyrme-Faddeev example, which we present here by way of illustrating the extrapolations in the subsequent cases in higher
dimensions.

Clearly, in this case the maximal and minimal levels of symmetry impositions coincide since there is only one $2$-plane in $\mathbb R^3$. Subject to this
symmetry, the most general Ansatz is
\be
\label{ansatz13}
Z=\left[
\begin{array}{c}
a+ib\\
c\,\e^{in\vf}
\end{array}
\right]\equiv
\left[
\begin{array}{c}
\sin\frac{f}{2}\,\e^{i\al}\\
\cos\frac{f}{2}\,\e^{in\vf}
\end{array}
\right]
\ee
where the functions $a$, $b$, $c$, $f$ and $\al$ all depend on both $\rho=\sqrt{|x_{\al}|^2}$
and $z\equiv x_3$, $\al=1,2$.

The Abelian Chern--Simons density \re{CS} on $\mathbb R^3$ is
\bea
\Omega_{\rm CS}^{(3)}&=&\vep_{mij}\,B_m\,G_{ij}\,.\label{CS3}
\eea
Substituting the azimuthally symmetric Ansatz \re{ansatz13} into \re{CS3} yields the simple expression
\be
\label{redCS3}
\Omega_{\rm CS}^{(3)}=-4\,\frac{n}{\rho}\,c\cdot\mbox{det}
\left|
\begin{array}{c}
a\ \ b\ \ c\\
\ a_{,\rho}\ b_{,\rho}\ c_{,\rho}\\
\ a_{,z}\ b_{,z}\ c_{,z}
\end{array}
\right|\,.
\ee
It is clear that if any one of the functions $a$, $b$, and $c$ vanishes, $\Omega_{\rm CS}^{(3)}$ vanishes.

Using the trigonometric parametrisation in the Ansatz \re{ansatz13}, $\Omega_{\rm CS}^{(3)}$ reduces to
\be
\label{CS3trig}
\Omega_{\rm CS}^{(3)}=\frac43\,\frac{n}{\rho}\left(F_{,\rho}\,\al_{,z}+{\rm antisymm.}(\rho,z)\right)\,,
\ee
where $F$ is the function
\be
\label{def1}
F(\rho,z)=\cos^3\frac{f}{2}.
\ee
The topological charge can then be expressed as,
\be
\label{topch1}
Q=\frac43\,2\pi\,\int\Omega_{\rm CS}^{(3)}\,\rho\,d\rho\,dz=\frac43\,2\pi\,n\,\int\,F_{[,\rho}\,\al_{,z]}\,d\rho\,dz
\ee
In analogy with the coordinates $\xi_{\mu}$ used in \re{xi4}-\re{xi5}, we denote the coordinates in the half plane $(\rho,z)=\xi_{\mu}$, $\mu=1,2$, $i.e.$,
\be
\label{xi1}
\xi_{\mu}=\left(
\begin{array}{c}
r\,\sin\psi\\
r\,\cos\psi
\end{array}\right)
\ee
with $0\le\psi\le\pi$, the volume integral \re{topch1} can be rewritten as follows
\bea
Q=\frac43\,2\pi\,n\,\int\,\vep_{\mu\nu}\,\pa_{i}F\,\pa_{\nu}\al\,d^2\xi&=&\frac43\,
2\pi\,n\,\int\,\left(F\,\hat x_i\,\vep_{\mu\nu}\,\,\pa_{\nu}\al\right)\big|_{r\to\infty}\,\hat\xi_{\mu}\,dS\nonumber\\
&=&\frac43\,2\pi\,\int_{\psi=0}^{\psi=\pi}F\,\pa_{\psi}\,\al|_{r\to\infty}\,\,d\psi\label{totdiv1}
\eea
where $\psi$ is the polar angle in the $(\rho,z)$ half plane.

Requiring the field configurations in question have the asymptotic values
\be
\label{asym}
\lim_{r\to\infty}f(r,\ta)=0\ ,\quad\lim_{r\to\infty}\al(r,\ta)=m\,\pi\,,
\ee
the integral \re{totdiv1} results in
\[
Q=\frac83\,n\,m\,\pi^2\,.
\]

\subsubsection{Chern-Simons charge on $\mathbb R^5$ subject to bi-azimuthal symmetry}
The Anstaz we use for the field \re{field} on $\R^5$ is
\be
\label{bi-az}
Z=\left[
\begin{array}{c}
a+ib\\
c_1\,\e^{in_1\vf}\\
c_2\,\e^{in_2\chi}
\end{array}
\right]\equiv\left[
\begin{array}{c}
\sin\frac12f\,\e^{i\al}\\
\cos\frac12f\,\sin g\,\e^{in_1\vf}\\
\cos\frac12f\,\cos g\,\e^{in_2\chi}
\end{array}
\right],
\ee
the functions $a,b,c_1,c_2$ depending
on the variables $\rho=\sqrt{|x_{\al}|^2}$, $\si=\sqrt{|x_{A}|^2}$ with $\al=1,2$,
$A=3,4$ and $z\equiv x_5$. $\vf$ and $\chi$ are the azimuthal angles in the $(x_1,x_2)$ and $(x_3,x_4)$
planes respectively, $(n_1,n_2)$ being the winding (vortex) numbers of these planes respectively.

The Abelian Chern--Simons density \re{CS} on $\mathbb R^5$ is
\bea
\label{CS5}
\Omega_{\rm CS}^{(5)}&=&\vep_{mijkl}\,B_m\,G_{ij}\,G_{kl}.
\eea
Substituting the bi-azimuthally symmetric Ansatz \re{bi-az} into \re{CS5} yields the simple expression
\be
\label{redCS5}
\Omega_{\rm CS}^{(5)}=32\,\frac{n_1n_2}{\rho\si}\,c_1\,c_2\cdot\mbox{det}
\left|
\begin{array}{cccc}
a\ \ \ b\ \ \ c_1\ \ \ c_2\\
\ a_{,\rho}\ b_{,\rho}\ c_{1,\rho}\ c_{2,\rho}\\
\ a_{,\si}\ b_{,\si}\ c_{1,\si}\ c_{2,\si}\\
\ a_{,z}\ b_{,z}\ c_{1,z}\ c_{2,z}
\end{array}
\right|\,.
\ee
Substituting the trigonometric parametrisation in \re{bi-az}, namely the parametrisation in which the sigma model
constraint is already imposed, \re{redCS5} simplifies to
\be
\label{redCS5trig}
\Omega_{\rm CS}^{(5)}=-\frac{n_1n_2}{\rho\si}\left[\pa_{\rho}F\,\pa_{\si}G\,\pa_z\al+{\rm cycl.\,symm.}(\rho,\si,z)\right]\,.
\ee
with the definitions
\be
\label{def2}
F(\rho,z)=\cos f+\frac{1}{2}\cos 2f\ ,\quad G=\cos 2g.
\ee
In analogy with the coordinates $\xi_{\mu}$ used in \re{xi4}-\re{xi5} and \re{xi1}, we denote the coordinates in the quarter sphere $(\rho,\si,z)=\xi_{\mu}$, ${\mu}=1,2,3$,
\be
\label{xi2}
\xi_{\mu}=\left(
\begin{array}{c}
r\,\sin\psi\,\sin\ta\\
r\,\sin\psi\,\cos\ta\\
r\,\cos\psi
\end{array}\right)
\ee
with $0\le\psi\le\pi$ and $0\le\ta\le\frac{\pi}{2}$, the volume integral of \re{redCS5trig}, namely the charge, can be
expressed as
\bea
Q&=&-(2\pi)^2\,n_1\,n_2\,\int\vep_{\mu\nu\la}\pa_{\mu}G\,\pa_{\nu}G\,\pa_{\la}\al\,d^3\xi\nonumber\\
&=&-(2\pi)^2\,n_1\,n_2\,\int\vep_{\mu\nu\la}\left(F\,\pa_{\nu}G\,\pa_{\la}\al\right)\bigg|_{r\to\infty}\,\hat\xi_{\mu}\,dS
\label{gauss}
\eea
in an obvious notation where $dS=r^2\,\sin\psi\,\ d\psi\,d\ta$, and where we have applied Gauss' Theorem.

The result is
\be
Q=4\,\pi^2\,n_1\,n_2\,\int_{\psi=0}^{\pi}\,\int_{\ta=0}^{\frac{\pi}{2}}\,F\,\left(
\pa_{\psi}G\,\pa_{\ta}\al-\pa_{\psi}\al\,\pa_{\ta}G\right)\bigg|_{r\to\infty}\,d\psi\,d\ta\,.
\ee
Requiring the field configurations in question have the asymptotic values
\be
\label{BCgal}
\lim_{r\to\infty}f=0\quad,\quad\lim_{r\to\infty}g=\ta\quad,\quad\lim_{r\to\infty}\al=m\,\pi\,,
\ee
\re{gauss} yields the following charge
\be
\label{charge}
\Omega_{\rm CS}^{(5)}=-12\,n_1\,n_2\,m\,\pi^3\,.
\ee

\subsubsection{Chern-Simons charge on $\mathbb R^7$ subject to tri-azimuthal symmetry}
The Anstaz we use is for the field \re{field} on $\mathbb R^7$ is
\be
\label{tri-az}
Z=\left[
\begin{array}{c}
a+ib\\
c_1\,\e^{in_1\vf}\\
c_2\,\e^{in_2\chi}\\
c_3\,\e^{in_3\xi}
\end{array}
\right]\equiv\left[
\begin{array}{c}
\sin\frac12f\,\e^{i\al}\\
\cos\frac12f\,\sin g\,\cos h\,\e^{in_1\vf}\\
\cos\frac12f\,\sin g\,\sin h\,\e^{in_2\chi}\\
\cos\frac12f\,\cos g\,\e^{in_3l\xi}
\end{array}
\right]
\ee
the functions $a,b,c_1,c_2,c_3$ depending
on the variables $\rho=\sqrt{|x_{\al}|^2}$, $\si=\sqrt{|x_{A}|^2}$, $\tau=\sqrt{|x_{a}|^2}$ with $\al=1,2$,
$A=3,4$, $a=5,6$ and $z\equiv x_7$. $\vf$, $\chi$ and $\xi$ are the azimuthal angles in the $(x_1,x_2)$, $(x_3,x_4)$
and $(x_5,x_6)$ planes respectively, $(n_1,n_2,n_3)$ being the winding (vortex) numbers of each plane respectively.

The Chern--Simons density \re{CS} on $\mathbb R^7$, is
\bea
\Omega_{\rm CS}^{(7)}&=&\vep_{pijklmn}\,B_p\,G_{ij}\,G_{kl}\,G_{mn}\label{CS7}
\eea
Substituting the tri-azimuthally symmetric Ansatz \re{tri-az} into \re{CS7} yields the simple expression
\be
\label{redCS7}
\Omega_{\rm CS}^{(7)}=96\,\frac{n_1\,n_2\,n_3}{\rho\si\tau}\,c_1\,c_2\,c_3\cdot\mbox{det}
\left|
\begin{array}{cccc}
a\, \ \ \ b\, \ \ \, c_1\ \ \, \ c_2\ \, \ c_3\\
a_{,\rho}\ \, b_{,\rho}\ \, c_{1,\rho}\ \, c_{2,\rho}\ \, c_{3,\rho}\\
a_{,\si}\ b_{,\si}\ c_{1,\si}\ c_{2,\si}\ c_{3,\si}\\
a_{,\tau}\ b_{,\tau}\ c_{1,\tau}\ c_{2,\tau}\ c_{3,\tau}\\
a_{,z}\ b_{,z}\ c_{1,z}\ c_{2,z}\ c_{3,z}
\end{array}
\right|\,.
\ee
Substituting the trigonometric parametrisation in \re{tri-az} ($i.e.$ the parametrisation in which the sigma model
constraint is already imposed) reduces the charge, namely the volume integral of \re{redCS7} to the simple expression
\be
\label{redCS7trig}
\Omega_{\rm CS}^{(7)}=-4\,\frac{n_1\,n_2\,n_3}{\rho\si\tau}\,\left[\pa_{\rho}F\,\pa_{\si}G\,\pa_{\tau}H\,\pa_z\al
+{\rm tot.\,antisymm.}(\rho,\si,\tau,z)\right]\,,
\ee
where
\be
\label{def3}
F(\rho,z)=\cos^6\frac{f}{2}\ ,\quad G=\frac14(1-\cos 2g)^2\ ,\quad H=\cos 2h.
\ee
In analogy with the coordinates $\xi_{\mu}$ used in \re{xi4}-\re{xi5} and \re{xi1}-\re{xi2}, we denote the coordinates in the sextant of the hypersphere
$(\rho,\si,\tau,z)=\xi_{\mu}$, $i=1,2,3,4$,
\be
\label{xi3}
\xi_{\mu}=\left(
\begin{array}{c}
r\,\sin\psi\,\sin\ta_1\,\sin\ta_2\\
r\,\sin\psi\,\sin\ta_1\,\cos\ta_2\\
r\,\sin\psi\,\cos\ta_1\\
r\,\cos\psi
\end{array}\right)
\ee
with $0\le\psi\le\pi$, \ $0\le\ta_1\le\frac{\pi}{2}$ and $0\le\ta_2\le\frac{\pi}{2}$,
the volume integral of \re{redCS7trig}, namely the charge, can be
expressed as
\bea
Q&=&-4\,(2\pi)^3\,n_1\,n_2\,n_3\,\int\vep_{\mu\nu\tau\la}\pa_{\mu}F\,\pa_{\nu}G\,\pa_{\tau}H\,\pa_{\la}\al\,
d^4\xi\nonumber\\
&=&-4\,(2\pi)^3\,n_1\,n_2\,n_3\,\int\vep_{\mu\nu\tau\la}\left(F\,\pa_{\nu}G\,\pa_{\tau}H\pa_{\la}\al\right)\bigg|_{r\to\infty}\,\hat\xi_{\mu}\,dS
\label{gauss1}
\eea
in an obvious notation where $dS=r^3\sin^2\psi\sin\ta_1\ d\psi\ d\ta_1\ d\ta_2$, and where we have applied Gauss' Theorem.
The result is
\be
Q=8\,\pi^3\,n_1\,n_2\,n_3\,\int_{\psi=0}^{\pi}\,\int_{\ta_1=0}^{\frac{\pi}{2}}\,\int_{\ta_2=0}^{\frac{\pi}{2}}\,F\,\left(
\pa_{\psi}G\,\,\pa_{\ta_1}H\,\pa_{\ta_2}\al+{\rm cycl.}(\psi,\ta_1,\ta_2)\right)\bigg|_{r\to\infty}\,d\psi\,d\ta_1\,d\ta_2\,.
\ee
Finally, requiring the boundary values
\be
\label{BCgal1}
\lim_{r\to\infty}g=\ta_1\ ,\quad\lim_{r\to\infty}h=\ta_2\  ,\quad\lim_{r\to\infty}\al=m\,\pi\,,
\ee
\re{gauss1} yields the following charge
\be
\label{charge1}
\Omega_{\rm CS}^{(5)}=64\,n_1\,n_2\,n_3\,m\,\pi^4\,.
\ee

\subsubsection{Chern-Simons charge on $\mathbb R^{2n+1}$ subject to $n$-fold azimuthal symmetry}
It is clear now that by induction, the appropriate imposition of symmetry is the application of azimuthal symmetry in each of the $n$ planes in
$\mathbb R^{2n+1}$. The resulting reduced subsystem will now be an $n+1$ dimensional system of PDE's, parametrised by $n+2$ functions
\[
a\,,\,b\,,\,c_1\,,\,c_2\,,\dots,c_n\ ,
\]
only $n+1$ of which are independent, subject to the sigma model constraint
\be
\label{constrn}
a^2+b^2+c_1^2+c_2^2+\dots +c_n^2=1\,.
\ee
The reduced Chern-Simons density will then take the form

\be
\label{redCS2n+1}
\Omega_{\rm CS}^{(2n+1)}\simeq\,\frac{n_1\,n_2\dots n_n}{\rho_1\rho_2\dots\rho_n}\,c_1\,\,c_2\,\dots c_n\cdot\mbox{det}
\left|
\begin{array}{cccccc}
a\ \ \ \ b\ \ \ \ c_1\ \ \ \, c_2\ \ .\ \ .\ \ c_n\\
a_{,\rho_1}\ b_{,\rho_1}\ c_{1,\rho_1}\ c_{2,\rho_1}\ .\ .\ c_{n,\rho_1}\\
a_{,\rho_2}\ b_{,\rho_2}\ c_{1,\rho_2}\ c_{2,\rho_2}\ .\ .\ c_{n,\rho_2}\\
\ .\ \ \ \ \ \, .\ \ \ \ \ \, .\ \ \ \ \ \, .\ \ \ \ \ \ .\ \ \ \, .\ \ \ \ \ \, .\\
\ .\ \ \ \ \ \, .\ \ \ \ \ \, .\ \ \ \ \ \, .\ \ \ \ \ \ .\ \ \ \, .\ \ \ \ \ \, .\\
a_{,\rho_n}\ b_{,\rho_n}\ c_{1,\rho_n}\ c_{2,\rho_n}\ .\ .\ c_{n,\rho_n}\\
a_{,z}\ \, b_{,z}\ \, c_{1,z}\ \, c_{2,z}\ .\ .\ c_{n,z}
\end{array}
\right|\,,
\ee
where of course $z=x_{n+1}$.

The quantity \re{redCS2n+1} is not as it stands explicitly a total divergence, but imposing the sigma model constraint \re{constrn} appropriately, it turns out to be ``essentially total divergence''. Alternatively, reparametrising it in terms of functions that satisfy the constraint,
$e.g.$ the polar parametrisations used above, it does become a total divergence explicitly.

The topological charge in this generic case is of the form
\be
\label{topgen}
Q\simeq n_1n_2\dots n_n\,m\,\pi^{n+1}\,,
\ee
where $n_1,n_2,\dots ,n_n$ are the vorticities in each of the $n$, $2$-planes.

\section{Hopfions in $D=5$ :  Numerical results for a restricted Ansatz}
The existence of a topological lower bound in itself does not guarantee that the model possesses nontrivial solutions. 
However, in the absence of exact solutions in closed form, the only recourse is to construct
the configurations which minimise the energy density functional, in this case (\ref{model12n+1}), {\it numerically}.

For $n=1$, $i.e.$ the $\mathbb C \mathbb P^1$ ``Skyrme-Faddeev'' models on $\mathbb R^3$, this problem has been approached by various authors.
Restricting to configurations within the Ansatz (\ref{ansatz13}), we mention here only the early work 
\cite{Gladikowski:1996mb},
\cite{Faddeev:1996zj},
\cite{Faddeev:1997pf},
where evidence has been presented for the existence of smooth minimum energy configurations in the static axially symmetric sector.
For the solutions with the lowest topological charge, the energy density is maximal at the origin and the energy density isosurfaces are
squashed spheres, while in other cases the axially symmetric solutions have toroidal structure
(see $e.g.$ \cite{Radu:2008pp} for more details together with an extensive list of relevant references).

On general grounds, we expect that the general model  (\ref{model12n+1}) would present solutions with rather similar properties
also in higher dimension $n>1$. However, the numerical construction of such solutions is much harder in this case.
In what follows, we report some  numerical results which can be viewed as an indication for the existence of 
finite energy Hopfions of a $\mathbb C \mathbb P^2$ ``Skyrme-Faddeev'' model on $\mathbb R^5$.
The study of such configuration has been performed for a simplified version of the general bi-azimuthally symmetric Ansatz (\ref{bi-az}),
\begin{eqnarray}
\label{bi-az1}
Z=\left[
\begin{array}{c}
a(\rho,\sigma,z)+ib(\rho,\sigma,z)\\
c_1(\rho,\sigma,z)\,\e^{in_1\vf}\\
c_2(\rho,\sigma,z)\,\e^{in_2\chi}
\end{array}
\right] 
\end{eqnarray}
with $\rho$ and $\sigma$ the radial variables in the $(x_1,x_2)$ and $(x_3,x_4)$ planes respectively, while $z\equiv x^5$.
In terms of these coordinates, the $D=5$  line element reads
\begin{eqnarray}
\label{m1}
ds^2=d\rho^2+\rho^2 d\varphi^2+d\sigma^2+\sigma^2 d\chi^2+dz^2.
\end{eqnarray}
The next step here is to consider the coordinate trasformation
\begin{eqnarray}
\label{coord-transf}
\rho=r \sin \psi \sin \theta,~\sigma=r \sin \psi \cos \theta,~~z=r \cos \psi
\end{eqnarray}
(with $0\leq r<\infty$, $0\leq \psi\leq \pi $, $0\leq \theta\leq \pi/2 $),
such that (\ref{m1}) becomes
\begin{eqnarray}
\label{m2}
ds^2=dr^2+r^2\left(
d\psi^2+\sin^2 \psi d\Omega_3^2 \right),
~~~{\rm with}~~d\Omega_3^2=d\theta^2+\sin^2\theta d\varphi^2+\cos^2 \theta d\chi^2.
\end{eqnarray} 

Then the functions $a,b,c_1,c_2$ in (\ref{bi-az1}) would depend on $r,\psi,\theta$.
Remarkably it turns out that 
\begin{eqnarray}
\label{new-ans}
a=a(r,\psi),~~b=b(r,\psi),~~c_1=c(r,\psi)\sin \theta,~~c_2=c(r,\psi)\cos \theta
\end{eqnarray} 
is a consistent truncation\footnote{This is similar to the factorization of the $\theta-$dependence on the $S^3$
sphere  employed in the scalar field Ans\"atze in 
\cite{Hartmann:2010pm},
\cite{Dias:2011at}.}
of the general model, provided that 
\begin{eqnarray}
\label{new-ans1}
n_1=n_2=1.
\end{eqnarray} 
This restrictive Ansatz greatly reduces the complexity of the system and makes the numerical construction of at least the lowest topological charge solutions,
possible.

In this approach, the problem reduced to solving a set of three partial differential equations with dependence on only two coordinates, 
for the functions
$a,b,c$ subject to the constraint
\begin{eqnarray}
\label{constr-1}
a^2+b^2+c^2=1.
\end{eqnarray}
As usual, these equations result by varying (\ref{model15}) 
$w.r.t.$ the functions $a,b,c$,
the constraint (\ref{constr-1}) being imposed by using the Lagrange multiplier method 
(see $e.g.$ \cite{Radu:2008pp}).
The boundary conditions satisfied by the functions $a,b$ and $c$ are
\begin{eqnarray}
\label{bc1}
&&
a\big|_{r=0}=-1,~~b\big|_{r=0}=0,~~c\big|_{r=0}=0,
~~a\big|_{r=\infty}=1,~~b\big|_{r=\infty}=0,~~c\big|_{r=\infty}=0,
\\
\nonumber
&&
\partial_\psi a\big|_{\psi=0,\pi}=0,~~\partial_\psi b\big|_{\psi=0,\pi}=0,~~c\big|_{\psi=0,\pi}=0.
\end{eqnarray}
We have restricted our considerations here to a particular truncation of 
the energy density funtional (\ref{model15}), with $\kappa_0=\kappa_1=\kappa_3=0$,
$i.e.$ a model consisting of the terms $G_{ij}^2$ and $G_{ijkl}^2$ only. 
This is the simplest model consistent with the Derrick scaling requirement for
finite energy.

Subject to this restrictive symmetry, these terms simplify to the expressions
\begin{eqnarray}
\label{G2-1}
G_{ij}^2=8
\left (
\frac{1}{r^2}(a_{,r}b_{,\psi}-b_{,r}a_{,\psi})^2
+\frac{c^2}{r^2\sin^2\psi}(  c_{,r}^2+\frac{1}{r^2}c_{,\psi})^2 
+\frac{c^4}{r^4\sin^4 \psi}
\right ),
\end{eqnarray}
and
\begin{eqnarray}
\label{G4-1}
G_{ijkl}^2=\frac{48c^4}{r^4\sin^4 \psi}\left( G_{ij}^2-\frac{8c^4}{r^4\sin^4\psi} \right),
\end{eqnarray}
with the action (which coincides with the total mass-energy):
\begin{eqnarray}
\label{action1}
E=2\pi^2 \int dr d\psi~ r^4\sin^3\psi \left(\frac{1}{4}\kappa_2^4 G_{ij}^2+ \frac{1}{16}\kappa_4^8 G_{ijkl}^2 \right).
\end{eqnarray}
The equations satisfied by  the functions $a,b,c$ are very complicated and are not presented here.

\begin{figure}
\begin{center}
\includegraphics{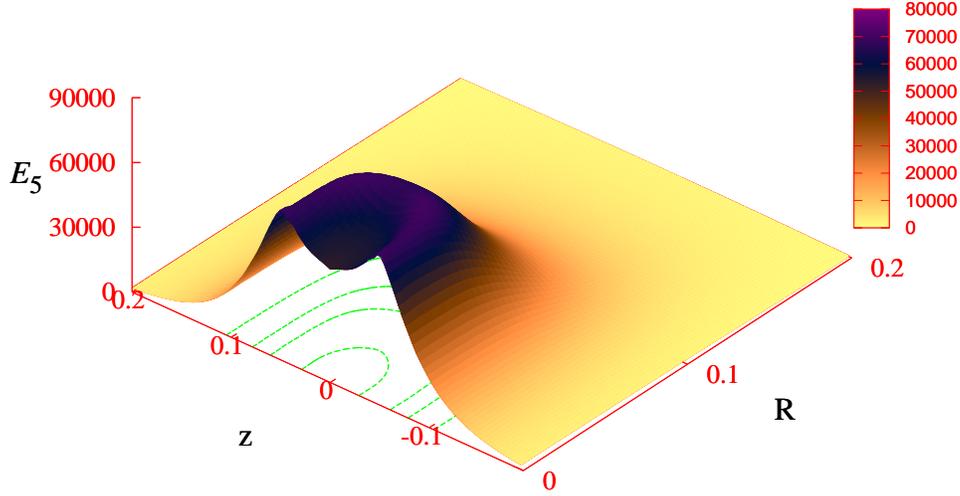}
\caption{
The energy density ${\cal E}_5$ is plotted for the $D=5$ Hopfion with $n_1=n_2=m=1$, 
as a function of
the coordinates $z=r\cos \psi$, $R=r \sin \psi$ .
}
\end{center}
\label{rod2}
\end{figure}

Our numerical approach is similar to that employed in \cite{Radu:2008pp} to construct various solutions, in particular axially symmetric Hopfions in $D=3$,
by directly solving the field equations. The numerical calculations were performed with the software 
package CADSOL, based on the Newton-Raphson method \cite{FIDISOL}. In this case, the field equations are first discretized on a 
nonequidistant grid and the resulting system is solved iteratively until convergence is achieved.
In this scheme, a new radial variable $x=r/(1+r)$ is introduced which maps the semi-infinite region $[0,\infty)$ to the closed region $[0,1]$.

Our numerical results strongly suggest the existence of a solution  of the model satisfying the boundary conditions (\ref{bc1}).
A distinguishing property of this configuration is that the function $a^2+b^2$ vahishes 
on a circle of finite radius\footnote{Note that any point there represents in fact a round three-sphere.} 
in the equatorial 'plane' $\psi=\pi/2$, such that $m=1$ in the asymptotic conditions (\ref{BCgal}).
The profile of the energy density of the solution is plotted\footnote{The data plotted 
in Figure 1 has been found for $\kappa_2^2=0.04$, $\kappa_4^8=1/24$, 
the total mass-energy of this configuration being $E=20.45$.
Although the model has no free parameters,
a choice of $\kappa_2$, $\kappa_4$ around these values
leads to
a faster convergence of our numerical algorithm.
However, we have verified that, within the numerical errors, the same results
are found for other choices of these parameters in the same range. 
} in Figure 1. 
One can see that  
the typical energy density isosurfaces are squashed spheres 
(this is similar to the $D=3$ solutions with the lowest topological charges \cite{Gladikowski:1996mb}).
However, a 'toroidal' shape is obtained when considering instead isosurfaces close to
the maximal value of the energy density ($e.g.$ ${\cal E}_5=7000$). 

Here we have to emphasize that given the complexity of, even the simplified system, the accuracy for the  configurations supplied by the solver 
is not really satisfactory.
For any grid choice, we could not reduce the typical numerical error (as supplied by the solver or based on the virial identity (\ref{virial2}))
below 10\%.
 However, 
based on the experience with related problems, we judge this to be 
a numerical problem mainly due to the extreme 
nonliniarity of the model and to the lack of  good starting profiles. 
As such, we interpret our results  only as an indication for the existence of the lowest charge
solution ($i.e.$ with $n_1=n_2=m=1$) of the $D=5$ model (\ref{model15}). 
We do, however, expect the qualitative features mentioned above to be confirmed by a more 
accurate numerical treatment of the problem.


\section{Summary and discussion}
We have proposed a family of sigma models on $\mathbb R^{2n+1}$ that may support knotted solitons solutions, and have established an energy lower bound for
these. The topological charge stabilising these solitons is the Chern-Simons charge, namely the volume integral of the Abelian Chern-Simons density,
in the given dimension. Like in the case of the Faddeev-Skyrme Hopfion, this lower bound is expressed as a fractional power of the topological charge.

The sigma models in question are the most general $\mathbb {CP}^n$ models on $\mathbb R^{2n+1}$ displaying a potential terms and all possible
(kinetic) ``Skyrme terms''. The latter
are the squares of $N$-forms constructed from the derivatives of the $\mathbb C \mathbb P^n$ fields, the highest of these being the $n$-form. (This contrasts with the
case of the usual Skyrmions on $\mathbb R^D$, where the highest order ``Skyrme term'' is the square of a $2D$-form, this being of $2$ orders higher than that of
the Hopfion.)

The criterion used in the choice of the sigma models is that the Chern-Simons density be expressed as a total divergence. This is so that the volume integral
of this density, which is the topological charge, be evaluated easily as a surface integral,
resulting in the transparent calculation of the topological charge. The Chern-Simons densities in question are constructed from the
composite Abelian connections (and their curvatures) of the $\mathbb C \mathbb P^n$ fields on $\mathbb R^{2n+1}$, and because of this we have chosen to refer to the Hopfions
pertaining to these models, as ``Abelian Hopfions''. These densities are manifestly total divergence when the
sigma model constraint is imposed. Otherwise, the are ``essentially total divergence'' in the sense 
that they yield no equations of motion when subjected to
the variational principle, with the constraint imposed. 
We note that this is precisely the situation with the Faddeev-Skyrme
Hopfion on $\mathbb R^3$. The latter is a $O(3)$ sigma model, which is is classically the same as the $\mathbb C \mathbb P^1$ model via the well known equivalence of these two
models.

The $\mathbb C \mathbb P^n$ systems considered are subjected to spatial symmetries. We have invoked the criterion that after imposition of symmetry, the residual
Chern-Simons density is expressed as a totally antisymmetrised product of the partial derivatives of the residual functions parametrising the system. This
enables the transparent imposition of suitable boundary values of the Hopfion field configurations, consistent with integer topological charge.
It turns out that the minimal symmetry necessary is that of one azimuthal (axial) symmetry, applied in any one $2$-plane in $\mathbb R^{2n+1}$.
We have also considered what we have
called the case of maximal imposition of symmetry, involving the application of azimuthal symmetry in each of the $n$, $2$-planes in $\mathbb R^{2n+1}$. We have
implemented both these types of symmetry and have stated the boundary conditions that must be satisfied for the Chern-Simons charge to be integer.
It is clearly possible to impose several intermediate levels of
symmetry in each case, the number of these increasing with $n$. These we have eschewed in the present work.

We have supplied an analogue of the classic analysis of Vakulenko and Kapitansky for the Faddeev-Skyrme Hopfion on $\mathbb R^3$, to establish an
energy lower bound on the Hopfions of the $\mathbb C \mathbb P^n$ models proposed here.
It is worth pointing out that the fractional-exponent topological lower
bound of the energy derived here is valid in all odd dimensions
which is in sharp contrast with that derived in all the Hopf dimensions,
$D=4n-1$, aimed at directly extending the Faddeev model in \cite{Lin-Yang12,Lin-Yang2}.
The reason for such a big distinction is that our topological
invariant here is the pure Chern-Simons invariant while that employed in \cite{Lin-Yang12,Lin-Yang2}
is the Hopf invariant which can only be defined in $4n-1$ dimensions. In
this sense, our work here complements that in \cite{Lin-Yang12,Lin-Yang2}, in that it
covers all odd dimensions.

As a final remark, we note that the cornerstone of our considerations is the definition of the topological charge, namely the volume integral of the
Chern-Simons density. The latter quantity is defined in terms of the composite connection (and curvature) of a nonlinear sigma model. In the present
work, we restricted our attention to Abelian such connections and curvatures, and it turned out that the $\mathbb C \mathbb P^n$ sigma models on $\mathbb R^{2n+1}$ were
precisely suited for this purpose. It is obvious that the same considerations can be applied to Chern-Simons densities defined in terms of non-Abelian
connections. In that case, the relevant sigma models are Grassmannian like systems. We shall report on such Hopfions elsewhere.

\bigskip
\bigskip
\noindent
{\bf\large Acknowledgements}\\
This work was started as part of project RFP07-PHY of Science Foundation
Ireland (SFI). E.R. gratefully acknowledge support by the DFG,
in particular, also within the DFG Research
Training Group 1620 ``Models of Gravity''. 
E.R. and D.H.T. would like to thank David Forster
for collaboration on the issue of numerical Hopfions
in $D=5$ dimensions, and are also grateful to Olaf Lechtenfeld, Yasha Shnir,
Paul Sutcliffe and especially Mikhail Volkov for valuable discussions.

\newpage

\begin{small}

\end{small}


\begin{thebibliography}{99}
\bibitem{F}
L.D. Faddeev,
{\it Knotted solitons,}
in Proc. Int. Congress Mathematicians, vol {\bf I}, pp. 235-244. Beijing, China: Higher Education Press (2002).
\bibitem{Faddeev:1996zj}
  L.~D.~Faddeev and A.~J.~Niemi,
  Nature {\bf 387} (1997) 58
  [hep-th/9610193].
\bibitem{Battye:1998pe}
  R.~A.~Battye and P.~M.~Sutcliffe,
  Phys.\ Rev.\ Lett.\  {\bf 81} (1998) 4798
  [hep-th/9808129].
\bibitem{Radu:2008pp}
  E.~Radu and M.~S.~Volkov,
  Phys.\ Rept.\  {\bf 468} (2008) 101
  [arXiv:0804.1357 [hep-th]].
\bibitem{Manton:2004tk}
  N.~S.~Manton and P.~Sutcliffe, {\it Topological solitons,}
  Cambridge, UK: Univ. Pr. (2004).
\bibitem{Tchrakian:2010ar}
  D.~H.~Tchrakian,
  J.\ Phys.\ A {\bf 44} (2011) 343001
  [arXiv:1009.3790 [hep-th]].
\bibitem{Vakulenko:1979uw}
  A.~F.~Vakulenko and L.~V.~Kapitansky,
  Sov.\ Phys.\ Dokl.\  {\bf 24} (1979) 433.
\bibitem{Lin-Yang1}
F. Lin and Y. Yang,
Commum. Math. Phys. {\bf 249} (2004) 273.
\bibitem{Deser:1982vy}
  S.~Deser, R.~Jackiw and S.~Templeton,
  Phys.\ Rev.\ Lett.\  {\bf 48} (1982) 975.
\bibitem{Lin-Yang12}
F. Lin and Y. Yang,
Nucl. Phys. B {\bf 747} (2006) 455.
\bibitem{Lin-Yang2}
F. Lin and Y. Yang,
Proc. R. Soc. A, {\bf 464} (2008) 2741.
\bibitem{Foster:2010zb}
  D.~Foster,
  Phys.\ Rev.\ D {\bf 83} (2011) 085026
  [arXiv:1012.2595 [hep-th]].
\bibitem{Gladikowski:1996mb}
  J.~Gladikowski  and M.~Hellmund,
  Phys.\ Rev.\ D {\bf 56} (1997) 5194
  [hep-th/9609035].
\bibitem{Faddeev:1997pf}
  L.~D.~Faddeev  and A.~J.~Niemi,
  hep-th/9705176.
\bibitem{Hartmann:2010pm}
  B.~Hartmann, B.~Kleihaus, J.~Kunz and M.~List,
  Phys.\ Rev.\ D {\bf 82} (2010) 084022
  [arXiv:1008.3137 [gr-qc]].
\bibitem{Dias:2011at}
  O.~J.~C.~Dias, G.~T.~Horowitz and J.~E.~Santos,
  JHEP {\bf 1107} (2011) 115
  [arXiv:1105.4167 [hep-th]].
\bibitem{FIDISOL}
W. Sch\"onauer and R. Wei\ss, J. Comput. Appl. Math. {\bf 27}, 279 
(1989);
\newline
M. Schauder, R. Wei\ss\ and W. Sch\"onauer, 
{\it The CADSOL Program Package},
 Universit\"at Karlsruhe, Interner Bericht Nr. 46/92 (1992);
\newline 
W. Sch\"onauer and E. Schnepf,  ACM Trans. on Math. Soft. {\bf13}, 
333 (1987).  


\end{thebibliography}
\end{document}